\documentclass[review]{elsarticle}

\usepackage{lineno,hyperref}

\usepackage{framed,multirow}

\usepackage{times}
\usepackage{epsfig}
\usepackage{graphicx}
\usepackage{pifont}
\usepackage{graphics}
\usepackage{printlen}
\usepackage{booktabs,subcaption,amsfonts,dcolumn}
\usepackage[dvipsnames]{xcolor}

\usepackage{amssymb}
\usepackage{latexsym}

\usepackage{url}
\usepackage{xcolor}

\usepackage{hyperref}
\usepackage{amsmath,amssymb,amsfonts}
\usepackage{algorithmic}
\usepackage{graphicx}
\usepackage{textcomp}
\usepackage{multirow}
\usepackage{caption}
\usepackage{url}
\usepackage{hyperref}
\usepackage{soul}
\usepackage{cleveref}
\usepackage{makecell}
\usepackage{xcolor}
\usepackage{tikz}
\usepackage{float}

\usepackage{latexsym}
\usepackage{rotating}
\usepackage{enumitem} 
\modulolinenumbers[5]
\journal{Journal of Medical Image Analysis}

\begin{document}

\begin{frontmatter}

\title{ Calibrate the inter-observer segmentation uncertainty via diagnosis-first principle }


\author[2]{Junde Wu}
\author[2]{Huihui Fang}
\author[2]{Haoyi Xiong}
\author[3]{Lixin Duan}
\author[4]{Mingkui Tan}
\author[5]{Weihua Yang}
\author[6]{Huiying Liu\corref{cor1}}
\author[2]{Yanwu Xu\corref{cor1}}  

\address[2]{Baidu Inc., Beijing, China}
\address[3]{School of Computer Science and Technology, University of Electronic Science and Technology of China, Chengdu, Sichuan, China}
\address[4]{School of Software Engineering, South China University of Technology, Guangzhou, Guangdong, China}
\address[5]{Big Data and Artificial Intelligence Institute, Shenzhen Eye Hospital, Jinan University, Shenzhen, Guangdong, China}
\address[6]{Institute for Infocomm Research, A*STAR, Singapore}
\cortext[cor1]{Corresponding authors: Huiying Liu (liuhy@i2r.a-star.edu.sg) and Yanwu Xu (ywxu@ieee.org).}

\begin{abstract}
    On the medical images, many of the tissues/lesions may be ambiguous. That is why the medical segmentation is typically annotated by a group of clinical experts to mitigate the personal bias. However, this clinical routine also brings new challenges to the application of machine learning algorithms. Without a definite ground-truth, it will be difficult to train and evaluate the deep learning models. 
    When the annotations are collected from different graders, a common choice is majority vote, e.g., taking the average of multiple labels. However such a strategy ignores the difference between the grader expertness. In this paper, we consider the task of predicting the segmentation with the calibrated inter-observer uncertainty. 
    We note that in clinical practice, the medical image segmentation is usually used to assist the \textbf{disease diagnosis}. Inspired by this observation, we propose diagnosis-first principle, which is to take disease diagnosis as the criterion to calibrate the inter-observer segmentation uncertainty. Following this idea, a framework named Diagnosis First segmentation Framework (DiFF) is proposed to estimate diagnosis-first segmentation from the raw images.
    Specifically, DiFF will first learn to fuse the multi-rater segmentation labels to a single ground-truth which could maximize the disease diagnosis performance. We dubbed the fused ground-truth as Diagnosis First Ground-truth (DF-GT).
    Then, we further propose Take and Give Model (T\&G Model) to segment DF-GT from the raw image.  
    In this way, DiFF can learn the segmentation with the calibrated uncertainty that facilitate the disease diagnosis. We verify the effectiveness of DiFF on three different medical segmentation tasks: optic-disc/optic-cup (OD/OC) segmentation on fundus images, thyroid nodule segmentation on ultrasound images, and skin lesion segmentation on dermoscopic images. 
    Experimental results show that the proposed DiFF is able to significantly facilitate the corresponding disease diagnosis (glaucoma diagnosis, thyroid cancer diagnosis and melanoma diagnosis, respectively), which outperforms previous state-of-the-art multi-rater learning methods.
\end{abstract}
\end{frontmatter}

\begin{figure}[t!]
    \centering
    \includegraphics[width=0.6\textwidth]{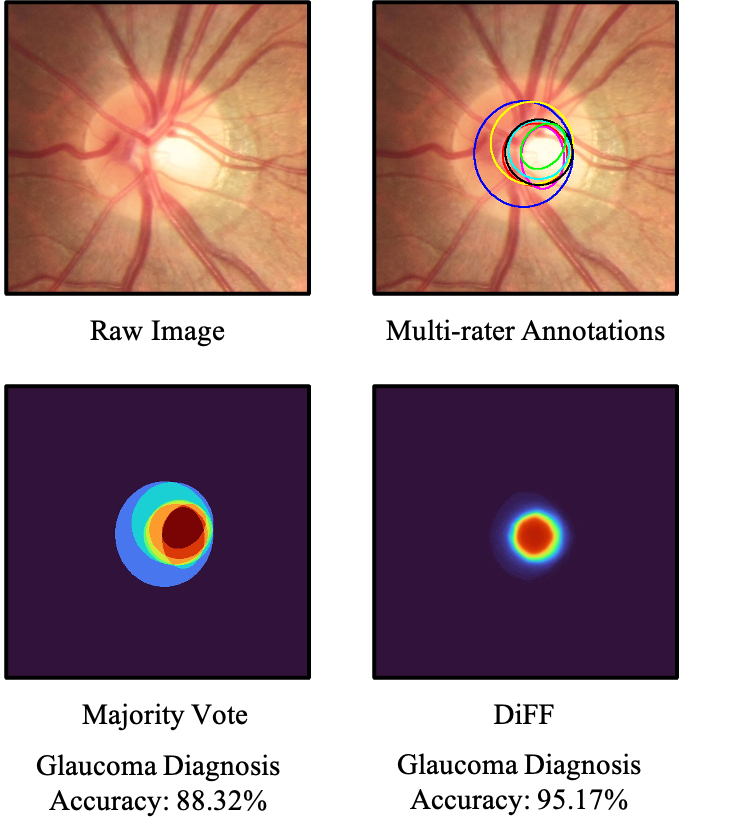}
    \caption{A multi-rater example of optic cup annotation. We can see that the inter-variance among the annotations is large. As shown in our experiments, the diagnosis results based on proposed DiFF are much more accurate than those of majority vote.}
    \label{fig:DiagSIM}
\end{figure}

\section{Introduction}
In most nature image segmentation tasks, the ground-truth of segmentation is unique and confident. However, in medical image segmentation, such an assumption is not always valid. It may be very expensive, or even infeasible to obtain the objective labels. The target tissues/lesions on medical images can be very ambiguous so that the labels collected are commonly subjective. 
Clinically, it is often necessary to collect the annotations from several different clinical experts to mitigate the individual bias. However, it comes at the cost of introducing inter-observer uncertainty to the annotations. The prior work (\cite{ji2021learning,yu2020difficulty,calhoun2009assessment}) called it 'multi-rater problem', which means that each instance of the dataset is annotated by several different raters. One multi-rater example of optic cup (OC) segmentation is shown in Figure \ref{fig:DiagSIM}. We can see that there has a significant variance in the annotations provided by different clinical experts. 
With such a large uncertainty on the segmentation label, it is not only hard to develop deep learning models as the automated segmentation solutions, but also can not quantitatively analyze and evaluate the models. 

When facing multi-rater problem, a simple and popular approach is to adopt majority vote, i.e., taking the average of multiple labels. Although this fusion strategy is simple and easy to implement, it comes at the cost of ignoring the different expertness of multiple graders (\cite{fu2020retrospective}). Recently, many works are proposed to model this inter-observer variability (\cite{guan2018said,chou2019every,kohl2018probabilistic,ji2021learning}) from the multi-rater labels. It is shown that a better modeling of the uncertainty will improve the final performance. However, they still need rater expertness provided to obtain a unique segmentation. Since rater expertness is commonly unavailable in the practice, in most cases, they are still limited to predict the traditional majority vote. 

Another branch of study proposed to jointly estimate the multi-rater expertness and the segmentation/classification (\cite{warfield2004simultaneous,raykar2010learning,albarqouni2016aggnet,cao2019max, wu2022learning}). These methods are generally based on Expectation–maximization (EM) algorithm, which the main idea is to use its own prediction to calibrate the ground-truth for the next-round supervision. However, the performance of these methods is largely depend on the segmentation/classification performance. When prediction itself is not good enough, it is likely to converge to an unsatisfied result.

We can see that we actually lack a gold criteria to evaluate the expertness of each rater, so as to calibrate the uncertainty in a correct way. We note that in clinical practice, medical image segmentation is commonly used for the disease diagnosis. Clinically, the disease diagnosis is usually conducted based on critical biomarkers derived from an analysis of the images. For example, on fundus images, the vertical Cup-to-Disc Ratio (vCDR) parameter computed from the optic cup/disc (OD/OC) masks is one of the most important clinical parameters for the glaucoma diagnosis (\cite{garway1998vertical}). In melanoma diagnosis, an asymmetrical and irregular shape of the skin lesions is a major biomarker indicating melanoma (\cite{gachon2005first}).
This inspires us to take disease diagnosis as a criterion to estimate the rater expertness and fuse the multi-rater labels. The fused label then can be used as the ground-truth for the segmentation training.

Specifically, we propose Diagnosis First Segmentation Framework (DiFF) to predict the calibrated segmentation masks based on the diagnosis-first principle. DiFF is implemented by two steps. The first step is to find the best fusion of the multi-rater labels for the diagnosis. In the first step, we use a diagnosis network to evaluate the multi-rater expertness. In particular, we optimize the expertness maps for each rater so as the fusion of multiple labels could maximize the diagnosis performance of the network. The label fused by these multi-rater expertness maps is named as Diagnosis First Ground-truth (DF-GT). In order to improve the generalization of DF-GT, we further propose Expertness Generator (ExpG) to generate the expertness maps in the optimization process. ExpG helps to constrain the high-frequency components by the continuity nature of neural network, so as to promise the generalization of DF-GT toward different diagnosis networks.
The second step is to segment DF-GT from the raw image. In the second step, we propose Take-and-Give Model (T\&G Model) to predict DF-GT from the raw images. T\&G Model is designed to integrate the diagnosis knowledge into the segmentation network, so as to better capture the segmentation features related to the diagnosis. The experiment shows that by adopting proposed DiFF, the estimated segmentation masks can significantly improve the performance of diagnosis, which outperforms SOTA multi-rater learning methods. 

In brief, the contributions of this paper can be summarized as follow:
\begin{itemize}
    \item To our knowledge, we are the first to address the multi-rater problem by using disease diagnosis as the criterion. According to the diagnosis-first principle, we propose a novel deep learning framework, called DiFF to calibrate the inter-observer uncertainty in the way to facilitate the disease diagnosis.
    \item As a part of DiFF, we propose a novel strategy to fuse the multi-rater labels. We evaluate the multi-rater expertness by a diagnosis network and fuse the labels according to the estimated expertness. The fused ground-truth is called DF-GT. In this process, we further propose ExpG to eliminate the high-frequency components by the continuity of neural network in order to improve the generalization.
    \item  We propose T\&G Model in DiFF to integrate the diagnosis knowledge into the segmentation network. T\&G Model can select and reinforce the diagnosis-related features by attentive feature interaction.
    \item We verify the proposed models on three different medical segmentation tasks. The experiment shows that the segmentation results gain high diagnosis performance, consistently outperform previous SOTA multi-rater learning methods.  
\end{itemize}

\section{Related Work}
\subsection{Medical Image Segmentation.} As an indispensable topic in medical image analysis, medical image segmentation has been researched by a large body of work. A majority of techniques used in natural image segmentation are also adopted to this task by previous work approaches (\cite{grow1,cluster1,mark1,alta1,def1}). With the advancement of parallel computation, neural network models are widely adopted to this task and outperform the traditional methods by a large margin (\cite{60,21,30,4,6}). Medical prior knowledge is commonly leveraged to improve neural network performance. For example, Fu et al. (\cite{fu2018disc}) segmented the optic disc and cup on polar transformed fundus images since optic disc and cup are naturally circle-like. Mirikharaji et al. (\cite{mirikharaji2018star}) encoded the star shape prior to the loss function for the segmentation of skin lesions. However, most existing methods assume that there has unique ground-truth label for each training instance and can not handle the uncertainty information.

\subsection{Learning with inter-observer uncertainty} With the supervised learning methods have been increasingly popular, the difficulty of obtaining objective and reliable labels is also magnified. Instead, in many cases, subjective labels from multiple experts are available. This raises the recent research attention (\cite{wdnet,19,2,24,42,ji2021learning}) on multi-rater problem. Many different methods are proposed to capture the uncertainty information from the multi-rater annotations (\cite{guan2018said,kohl2018probabilistic,baumgartner2019phiseg,ji2021learning,jensen2019improving}). However, they still need the users to provide the rater expertness in advance, which limits its application.

Another branch of study worked on estimating the rater expertness so as to fuse the multi-rater labels in a better way (\cite{warfield2004simultaneous,raykar2010learning,tanno2019learning}). Most of them relied on structural information of raw image as the prior to evaluate the multi-rater expertness. For example, Rayker \MakeLowercase{\textit{et al.}} (\cite{raykar2010learning}) learned to jointly predict the results and evaluate the rater expertness based on EM algorithm. Many works proposed after then following this idea (\cite{rodrigues2018deep,albarqouni2016aggnet,cao2019max}). Rodrigues \MakeLowercase{\textit{et al.}} (\cite{rodrigues2018deep}) extended this idea to neural-network-based classifiers to facilitate the learning. Cao \MakeLowercase{\textit{et al.}} (\cite{cao2019max}) adopted f-mutual information to evaluate the rater expertness. However, these methods cannot perform well when the source raw image is ambiguous, since they rely on the raw object structure to evaluate the multi-rater expertness. In addition, although most of these works have been done in the medical imaging scenario, medical prior knowledge is rarely adopted to address the problem. Thus there has no guarantee that these estimated labels will be helpful in clinical practice.




\section{Method}
\subsection{Motivation}\label{sec:motivation}
Clinically, the segmentation of lesions/tissues can significantly facilitate the disease diagnosis. The same thing can also be observed in deep learning based automated diagnosis models. Many prior works have shown the segmentation information can bring diagnosis networks with solid improvement (\cite{fu2018disc,zhou2019collaborative,wu2020leveraging,li2019attention,bajwa2019two,wu2022seatrans}). The common practices include the input concatenation, region of interest (ROI) extraction (\cite{fu2018disc,bajwa2019two}), channel attention  (\cite{li2019attention,zhou2019collaborative}), vision transformer based feature fusion (\cite{wu2022seatrans}), and transfer learning (\cite{wu2020leveraging}). However, when the segmentation contains underlying inter-observer uncertainty, the relationship between them comes to be more complicated. The 
segmentation of each rater may have different effects on the disease diagnosis. Thus, fusing the multi-rater segmentation in different ways will also differently affect the diagnosis performance. In order to quantitatively analyze these affects on the diagnosis, we perform a preliminary experiment with a OD/OC segmentation setting on the REFUGE-2 benchmark (\cite{fang2022refuge2}).

\begin{figure}[b]
    \centering
    \includegraphics[width=\textwidth]{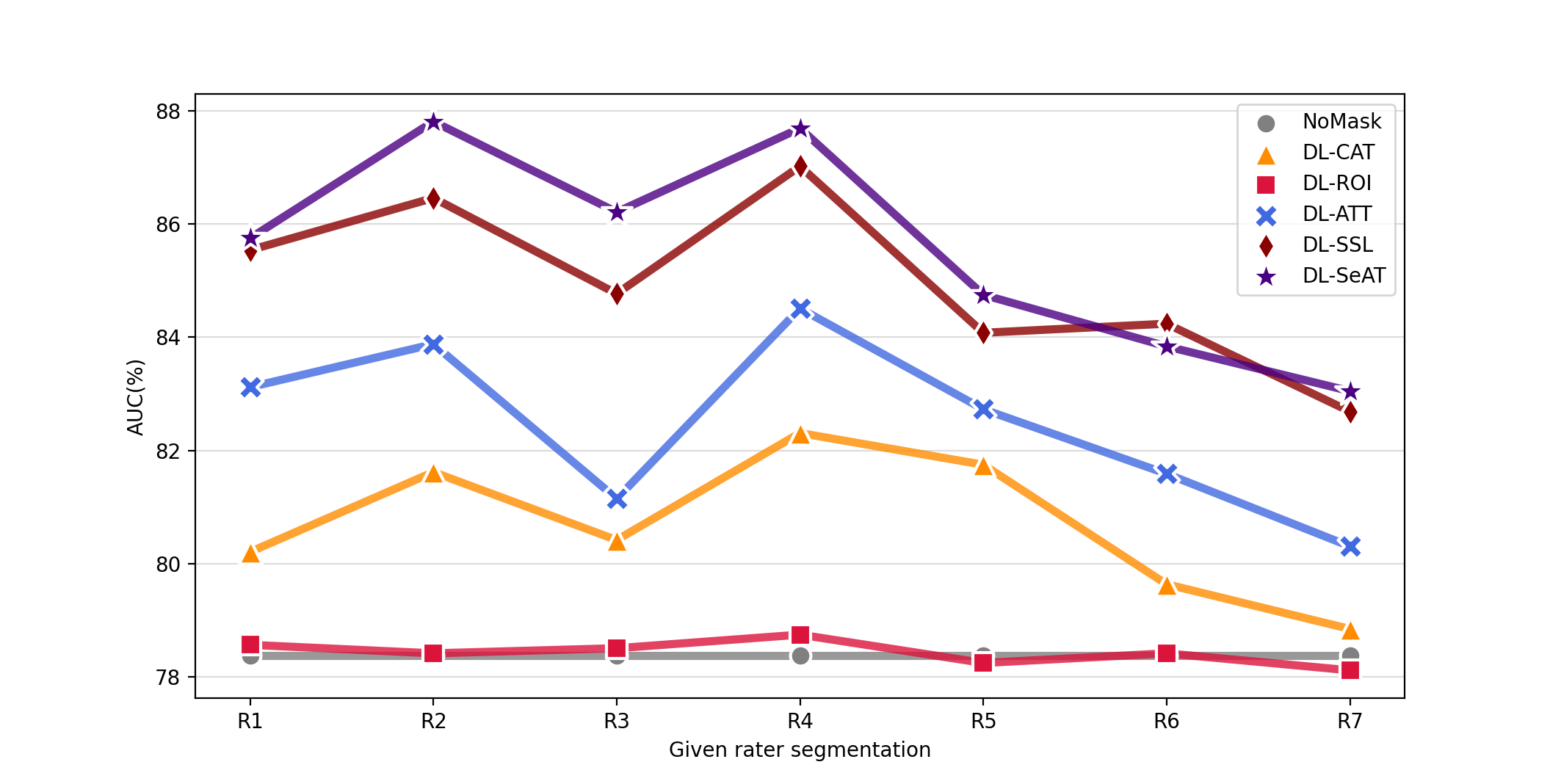}
    \caption{Diagnosis performance of various segmentation-assisted diagnosis models given the multi-rater segmentation. R1, R2, R3, etc denote the segmentation is annotated by rater1, rater2, rater3, etc. NoMask denotes predicting from only raw images.}
    \label{fig:motivation}
\end{figure}

We train aforementioned segmentation-assisted diagnosis networks, which are denoted as DL-CAT (\cite{d2021convit}), DL-ROI (\cite{bajwa2019two}), DL-ATT (\cite{li2019attention}), DL-SSL (\cite{wu2020leveraging}), and DL-SeAT (\cite{wu2022seatrans}) to diagnose glaucoma from the fundus images, with optic OD/OC segmentation masks as the auxiliary information. The network is trained to predict the diagnosis labels, which 0 refers to nonglaucoma and 1 refers to glaucoma. We then provide the models with different segmentation masks annotated by seven different raters respectively. We use ResNet50 as the diagnosis backbone in all these methods for a fair comparison. The final diagnosis performance measured by AUC score (\%) is shown in Figure \ref{fig:motivation}. 

Figure \ref{fig:motivation} shows that different rater's segmentation will have different effects on the diagnosis. It is clearly to see that Rater2 (R2) and Rater 4 (R4) improve diagnosis performance more than the other raters, and R7 brings the fewest improvements comparing with the others. Such a general pattern can be consistently observed on all kinds of segmentation-assisted diagnosis methods, except DL-ROI, which takes no significant improvement comparing with NoMask. These results demonstrate that some raters segmented the images with particular patterns which can be more beneficial to the diagnosis. These specific preferences can also be modeled by the neural networks, which is in line with the previous findings (\cite{ji2021learning}). 
Based on these observations, it can be inferred that training a segmentation network which can recognize and enhance the diagnosis-first rater preferences will help to calibrate the uncertainty in a way conducive to the disease diagnosis.


\subsection{DiFF}
Inspired by the previous experiment, we intend to find a specific fusion of the multi-rater segmentation labels which is able to maximize the diagnosis performance. By training the segmentation network on such a kind of ground-truth, we can learn to predict from the raw images the diagnosis-first segmentation masks. These masks will be helpful 
to the disease diagnosis.

\begin{figure*}[h]
    \centering
    \includegraphics[width=0.85 \textwidth]{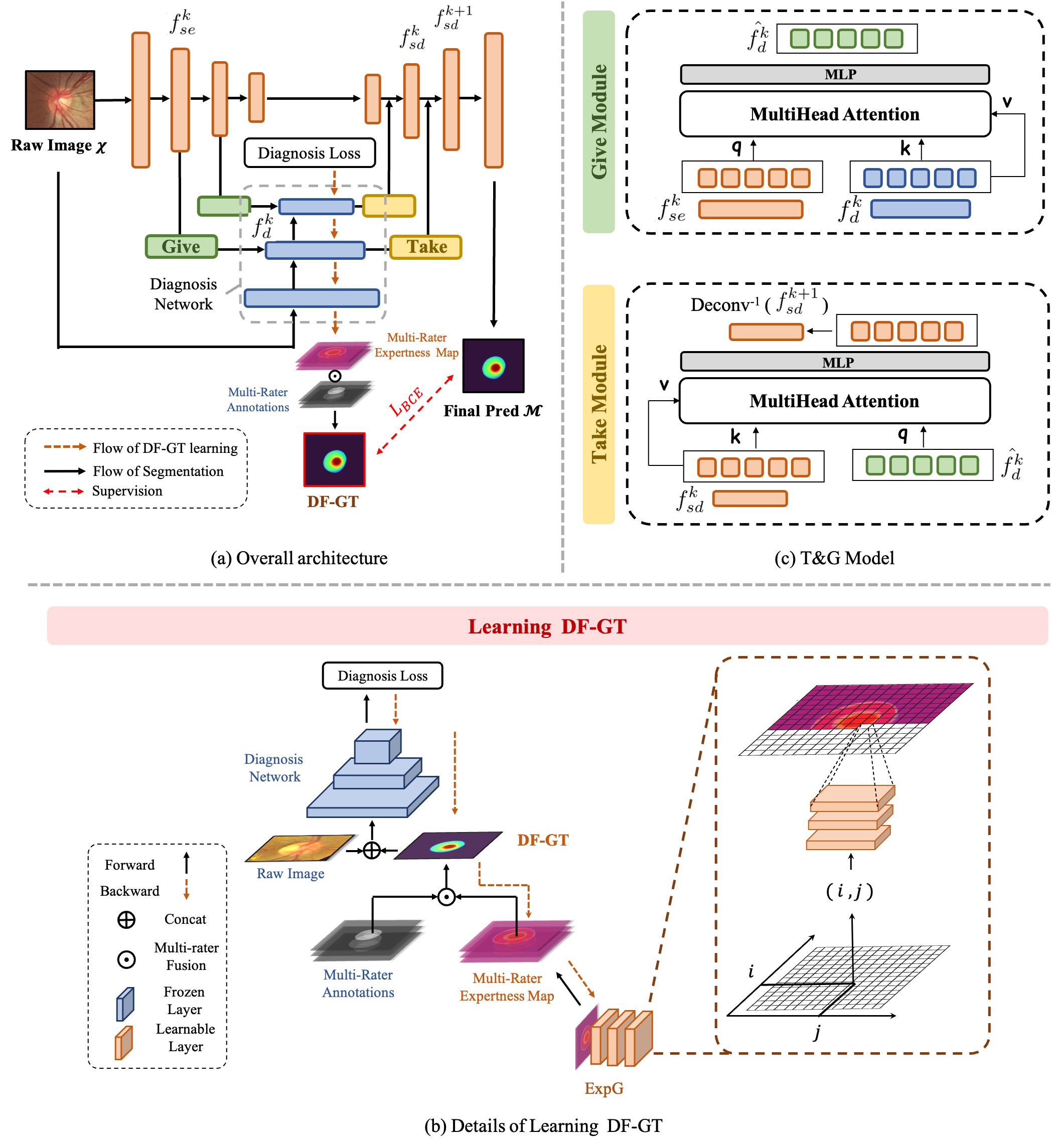}
    \caption{An illustration of the proposed method using OD/OC segmentation on fundus images as the example, which starts from (a) an overview of DiFF, and continues with zoomed-in diagrams of individual Models, including (b) the DF-GT learning flow, and (c) T\&G Model.}
    \label{fig:overall}
\end{figure*}


Following this idea, we propose DiFF, which the overall pipeline is shown in Figure \ref{fig:overall} (a). A weighted fusion of multi-rater segmentation labels, called DF-GT is first learned from a diagnosis network, as shown in Figure \ref{fig:overall} (b). DF-GT is then used as the ground-truth to train the segmentation network. In the segmentation network, the raw image $x$ is first sent to the segmentation encoder and diagnosis network to get the segmentation features and diagnosis features.
T\&G Model then connects the features in the segmentation encoder, the diagnosis network, and the segmentation decoder, to instill the diagnosis prior knowledge into the segmentation network. The estimated mask $\mathcal{M}$ of the segmentation decoder serves as the final prediction of the model. In the training stage, the model is end-to-end supervised by DF-GT using the binary cross entropy (BCE) loss:
\begin{equation}\label{equation:fd}
    \mathcal{L} = L_{bce}(\textrm{DF-GT}, \mathcal{M}).
\end{equation}
In the following sections, we introduce DF-GT learning process and T\&G Model in detail.

\subsubsection{Learning DF-GT}
In the running of DiFF, we first fuse the multi-rater labels to a unique ground-truth with the highest diagnosis performance. To fuse multiple labels to one ground-truth, we weighted sum the multiple labels by the expertness of each rater:
\begin{equation}\label{eqn:fusion}
   \textrm{Groundtruth} = s \odot m  = \sum_{i=1}^{n} s_{i} * m_{i}, 
\end{equation}
where $*$ denotes the element-wise multiplication, $s_{i}$ and $m_{i}$ are the annotation and the expertness map of the rater $i$, respectively.

We then assess the rater expertness through the disease diagnosis. In other words, the rater contributes more to the correct diagnosis would be given higher expertness. Toward that end, we take the multi-rater expertness maps as the learnable variables to maximize the diagnosis performance of a segmentation-assisted diagnosis network. The network can be implemented in any way. For the convenience, we use basic DL-CAT as an example, i.e., a standard classification network inputted by a concatenation of the raw image and the segmentation mask. 
We pre-train the network on the training set of the dataset to correctly evaluate the diagnosis. 

The overflow of learning DF-GT is shown in Figure \ref{fig:overall} (b). Formally, consider each raw image $x \in \mathbb{R}^{h\times w\times c}$ is annotated by $n$ raters, resulting in $n$ segmentation masks $s \in \mathbb{R}^{h\times w \times n}$, and the image is diagnosed as $y \in [0,1]$. Let $\theta$ denotes the set of diagnosis model parameters. $L (\theta, x, y)$ denotes the loss function of a standard classification task. Our goal is to find an optimal expertness map $m \in \mathbb{R}^{h\times w\times n}$ by solving the following optimization problem:
\begin{equation}\label{first order backer}
    m^{*} = \mathop{\arg\min}\limits_{m} L (\theta,x \oplus [s \odot m] ,y), 
\end{equation}
where $\oplus$ denotes concatenation operation, $m^{*}$ denotes the optimal expertness maps. 
Note that we applied Softmax on $m$ to normalize the weights. According to Eqn. (\ref{first order backer}), we are actually finding the expertness maps which can minimize the diagnosis loss. We name the fused ground-truth under these expertness maps, i.e., $s \odot m^{*}$ as DF-GT. We adopt gradient descent to solve Eqn. (\ref{first order backer}):
\begin{equation}\label{gradient}
    m^{t+1} = m^{t} + \alpha \nabla_{m} L (\theta,x \oplus [s \odot m] ,y),
\end{equation}
where $\alpha$ is the learning rate. 

\begin{figure}[h]
    \centering
    \includegraphics[width=0.7\textwidth]{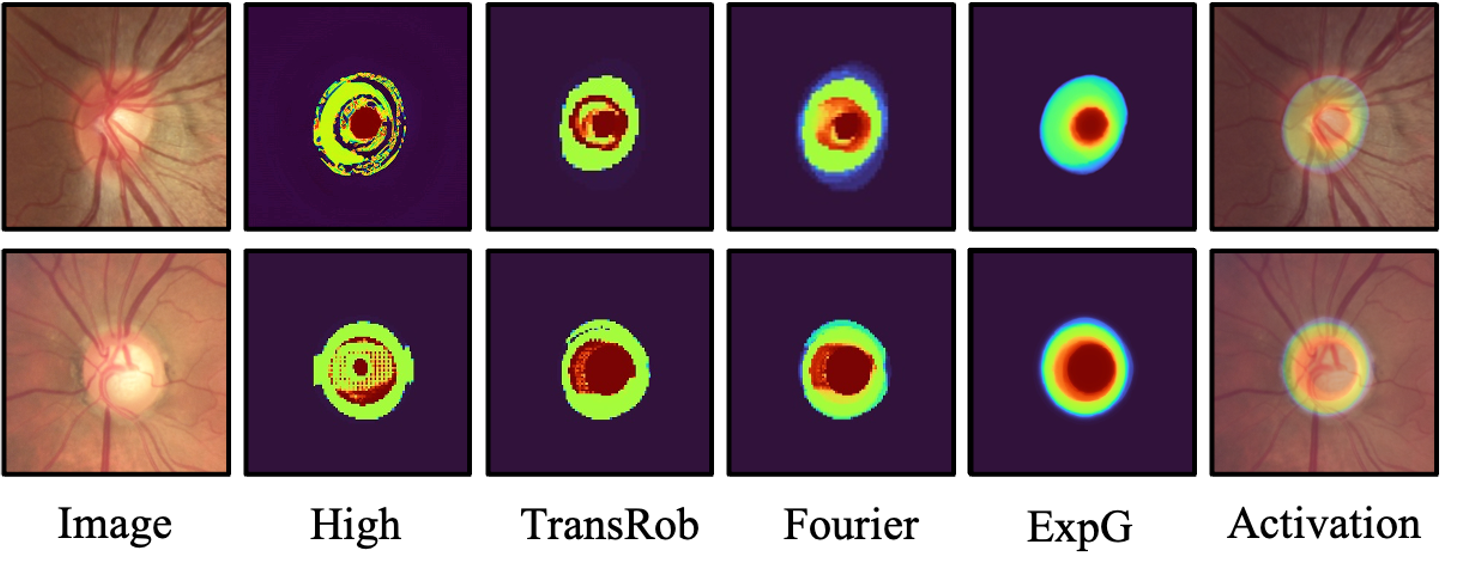}
    \caption{Visualized results of high-frequency elimination. It can be seen that ExpG generated DF-GT are smoother and more reasonable. }
    \label{fig:supvis}
\end{figure}

The results optimized in this way can improve the diagnosis performance to a high level. But since it is optimized toward one specific diagnosis network, it will not be general enough. The visualization results also show it suffers heavily from high-frequency noises. An example on OD/OC segmentation is shown in Figure \ref{fig:supvis}. The latest findings suggest that these high-frequency components have close relationship with the generalization capability (\cite{Wu2019UniversalTA,h3}). More specifically, high-frequency components are likely generated by the repeated grid effect of the transposed convolution  (\cite{warfield2004simultaneous}) when we backprop gradients through each convolution layer. Thus they will be very specific to the network architecture, like the number of the layers, the stride of the convolution, etc. 

Therefore, a possible approach to improve generalization of DF-GT is to constrain the high-frequency components in the optimization process. We tried several methods to achieve the goal, including Transformation Robustness (TransRob), Fourier Transform (Fourier) and proposed Expertness Generator (ExpG). Among them, Transformation Robustness constrains high-frequency gradients by applying small transformations to the expertness map before optimization. In practice, we rotate, scale and jitter the maps. Fourier Transform transforms the expertness map parameters to frequency domain, thus decorrelated the relationship between the neighbour pixels. 

Unlike the aforementioned non-learnable methods, the proposed ExpG constrains the high-frequency components by the continuity nature of the neural network, which is implemented by a tiny CNN based pixel generator (which consists four CNN layers in our implementation). An illustration of ExpG is shown in Figure \ref{fig:overall} (b). The network scans one pixel at a time. For each pixel it predicts the pixel value given the position of the pixel. The input of ExpG is the coordinate vector $ (i,j)$ and the output is the pixel value. ExpG is optimized to generate expertness maps which are able to minimize the diagnosis loss. Denote the parameters of ExpG as $\phi$, our goal is to solve:
\begin{equation}\label{eqn:lhc}
    \mathop{\arg\min}\limits_{\phi} L (\theta,x \oplus [s \odot \{\rm{ExpG}_{\phi} (i,j)\}_{i = 1 \sim h}^{j = 1 \sim w}] ,y),
\end{equation}
where $\{\rm{ExpG}_{\phi} (i,j)\}_{i = 1 \sim h}^{j = 1 \sim w}$ denotes generated expertness map with size $h \times w \times n$ by ExpG. Since the continuity of the neural network mapping function, similar inputs tend to cause similar outputs, which lead the element values in expertness maps variant smoothly between the positions and thus eliminate the high-frequency components. 

The visualized example of all high-frequency elimination methods are shown in Figure \ref{fig:supvis}. The first line is a non-glaucoma example and the second line is a glaucoma example. We can see the visualized effects of TransRob and Fourier are obviously improved comparing with high-frequency examples. But ExpG achieves much better visualized effects than both of them. In the experiments, we adopt ExpG to eliminate the high-frequency components in DF-GT optimization process. 

\subsubsection{T \& G Model}
In order to predict diagnosis-first ground-truth (DF-GT) from the given raw images for the inference, we need to train a segmentation network under its supervision. However, the standard segmentation network will show unsatisfied performance on DF-GT. Because  
first, the standard segmentation network absents diagnosis information. In DF-GT, the expertness is learned from the diagnosis network. The segmentation network that absents this prior knowledge is hard to capture the features related to the diagnosis, thus fail to learn DF-GT. Meanwhile, a fusion with fixed rater expertness will be easier to learn, since the static rater preference can be modeled by neural network (\cite{ji2021learning}). However, DF-GT is fused by dynamically generated expertness maps. In other words, the expertness maps of each sample are different. The standard segmentation network that lack dynamic adaptability may become hard to follow.

Towards the efficient learning of DF-GT, we propose Take and Give Model (T\&G Model) to dynamically instill the diagnosis features into the segmentation network. The architecture of T\&G Model is shown in Figure \ref{fig:overall} (c).
In particular, T\&G Model contains a Take Module and a Give Module. Give Module bridges the segmentation encoder and the diagnosis network, to fill the segmentation information into the diagnosis features. Through the attention mechanism, Give Module weights the diagnosis feature based on its affinity with the segmentation feature. Only the diagnosis features closely related to the segmentation will be selected and transformed, resulting in a transformed diagnosis feature that can be used for segmentation. Take Module bridges the transformed diagnosis feature and the segmentation decoder, to provide the diagnosis knowledge to the segmentation. Take Module is symmetrical to Give Module, but picks and transforms the segmentation features based on the diagnosis features. Take Module and Give Module will be applied on several corresponding layers of the segmentation encoder, the diagnosis network, and the segmentation decoder. We studied which layers should be connected in the ablation study. 

Specifically, T\&G Model is constructed by attention mechanism (\cite{vaswani2017attention}) following MLP layer. Consider T\&G Model at the $k^{th}$ layer. The inputs of Give Module are a segmentation feature map and a diagnosis feature map, which are the $k^{th}$ layers of segmentation encoder and diagnose network, respectively. The output of the module is a transformed diagnosis feature map with the same size as the input feature map. To deal with the feature map efficiently, we reshape the feature map $F \in \mathbb{R}^{H\times W\times C}$ into a sequence of flattened patches $f \in \mathbb{R}^{N\times (P^{2}\cdot C)}$, where $(H,W)$ is the resolution of the original feature map, $C$ is the number of channels, $(P, P)$ is the resolution of the patches, and $N = HW / P^{2}$ is the resulting number of patches. Consider the encoder segmentation feature map patches is $f_{se}^{k}$ and the diagnosis feature map patches is $f_{d}^{k}$, then Give Module can be represented as:
\begin{equation}\label{equation:fd}
    \hat{f_{d}^{k}} = {\rm MLP}({\rm Attention}(f_{se}^{k} + E_{se}^{k}, f_{d}^{k} + E_{d}^{k}, f_{d}^{k})),
\end{equation}
where $\hat{f_{d}^{k}}\in\mathbb{R}^{N\times (P^{2}\cdot C)}$ is the transformed diagnosis feature, ${\rm Attention}(query, key, value)$ denotes multi-head attention mechanism, MLP is multi-layer perceptron, $E_{se}^{k}, E_{d}^{k} \in \mathbb{R}^{N\times (P^{2}\cdot C)}$ are positional encodings (\cite{carion2020end}) for the segmentation feature map and diagnosis feature map respectively. A standard attention mechanism is adopted here, which is to first calculate an affinity weight map $a \in \mathbb{R}^{N \times (H \cdot W)}$ between the $query$ and $key$, and then use it to select the features in $value$. The affinity weight map $a$ is represented as:
\begin{equation}
    a = {\rm softmax}(\frac{q(f_{se}^{k} + E_{se}^{k})k(f_{d}^{k} + E_{d}^{k})^{T}}{\sqrt{P^{2}\cdot C}}),
\end{equation}
where the functions $q(\cdot)$, $k(\cdot)$ denote the linear mappings for the inputs of $query$ and $key$. We can see the normalized affinity weights $a$ are defined by how each diagnosis feature is influenced by all the segmentation features. These weights are then applied to all the diagnosis features in $value$,
\begin{equation}
    {\rm Attention}(f_{se}^{k} + E_{se}^{k}, f_{d}^{k} + E_{d}^{k}, f_{d}^{k}) = a \cdot v(f_{d}^{k}),
\end{equation}
where $v(\cdot)$ is the linear mapping for $value$.

We then apply a Multi-layer Perceptron (MLP) to further select the reinforced attention result. MLP has two linear mappings with weight $W_{f1}, W_{f2} \in \mathbb{R}^{(P^{2}\cdot C) \times (P^{2}\cdot C)}$ and a GELU (\cite{hendrycks2016gaussian}) activation function in its module:
\begin{equation}
    {\rm MLP}(f) = {\rm GELU}(f \cdot W_{f1}) \cdot W_{f2}.
\end{equation}

Take Module provides the segmentation decoder with the diagnosis knowledge, to get a diagnosis-focused segmentation result. Take Module is an attention-based module symmetrical to Give Module, which conversely uses combined feature $\hat{f_{d}}$ as $query$ and segmentation feature $f_{sd}$ as $key$ and $value$, so as to select and transform the segmentation features based on the diagnosis knowledge. Take Module is applied on the features before the standard deconvolution layer in the decoder. Consider the $k^{th}$ layer in segmentation decoder is $f_{sd}^{k}$, then Take Module interact with decoder by:
\begin{equation}\label{equation:fd}
    f_{sd}^{k+1} = {\rm Deconv} ({\rm MLP}({\rm Attention}(\hat{f_{d}^{k}} + E_{d}^{k}, f_{sd}^{k} + E_{sd}^{k}, f_{sd}^{k}))).
\end{equation}

Take Module selects the diagnosis features by the segmentation features in the segmentation encoder, and Give Module enhances the diagnosis related segmentation features in the segmentation decoder. Through this way, the segmentation network is able to capture the diagnosis calibrated segmentation features in DF-GT, so as to efficiently predict DF-GT from the raw images in the inference stage.

\section{Experiment}
\subsection{Tasks and Datasets}
Extensive experiments are conducted to verify the effectiveness of the proposed framework. We conduct the experiments on three different medical segmentation tasks with data from varied image modalities, including color fundus images, ultrasound images and dermoscopic images.

\textbf{OD/OC Segmentation \& Glaucoma Diagnosis} On the fundus images, we calibrate the OD/OC segmentation via glaucoma diagnosis. The experiments are conducted on REFUGE-2 benchmark (\cite{fang2022refuge2}), which is a publicly available dataset for OD/OC segmentation and glaucoma classification. It contains in total 1200 color fundus images, including three sets each with 400 images for training, validation, and testing. Seven glaucoma experts from different organizations labeled the OD/OC contour masks manually for the REFUGE-2 benchmark. 120 samples correspond to glaucomatous subjects, and the others correspond to non-glaucomatous subjects. The glaucomatous subjects are distributed equally to the training, validation, and test set.

\textbf{Thyroid nodules Segmentation \& Thyroid cancer Diagnosis} On ultrasound images, we calibrate thyroid nodules segmentation via thyroid cancer diagnosis. The experiments are conducted on TNMIX benchmark, which is a publicly available mixed dataset for thyroid nodule segmentation and diagnosis. It contains in total 5191 ultrasound images from two sources, including 4554 images from TNSCUI (\cite{tnscui}) and 637 images from DDTI (\cite{ddti}). The images are segmented manually by five graders, following the same annotation protocol, and their annotations were approved by experienced thyroid-radiologists. The images are diagnosed as malignant or benign thyroid nodule, where 2881 samples correspond to malignant subjects, and the others correspond to benign subjects. 2732 samples are used for the training, 911 for the validation and the rest 911 for testing. The proportion of malignant samples in each set is made roughly the same.

\textbf{Skin lesions Segmentation \& Melanoma Diagnosis} On dermoscopic images, we calibrate skin lesions segmentation via melanoma diagnosis. The experiments are conducted on ISIC (\cite{codella2019skin}), which is an open source dataset of skin lesions segmentation and diagnosis. The images are associated with ground-truth diagnosis of melanoma and skin lesions masks. 1600 images with multiple skin lesions masks are selected to conduct the experiment. Skin lesions are annotated individually by four recognized skin cancer experts. 312 samples correspond to melanoma subjects, and the others correspond to non-melanoma subjects. The dataset is split to three sets with 960 images for training, 320 for validation and 320 for testing. The proportion of malignant samples in each set is made roughly the same.



\subsection{Experimental Setup}
\subsubsection{Implementation Details}
To verify the proposed method, we train two baselines for diagnosis and segmentation, denoted as Baseline-Dia and Baseline-Seg, respectively. Baseline-Dia is implemented by a segmentation attentive diagnosis network (\cite{zhou2019collaborative}) using ResNet50 (\cite{resnet}) as the backbone. In the pre-training stage, we train the network with majority vote segmentation masks on the training set.
Baseline-Seg is a standard ResUnet (\cite{yu2019robust}) using ResNet50 as the backbone. In DiFF, we use the frozen pre-trained Baseline-Dia as the diagnosis network connected by T\&G Model. Pre-training is preformed on the training set of the selected dataset.

All the experiments are implemented with the PyTorch platform and trained/tested on 4 Tesla P40 GPU with 24GB of memory. All training and test images are uniformly resized to the dimension of 256$\times$256 pixels. DF-GT is trained using Adam optimizer (\cite{kingma2014adam}) for 125 epochs. DiFF is trained end-to-end using Adam optimizer with a mini-batch of 16 for 80 epochs. The learning rate is always set to 1 $\times 10^{-4}$. The detailed network structures and hyper-parameters can be found in our released code.
\subsubsection{Evaluation Metric}
The diagnosis performance is evaluated by AUC (Area Under the receiver operating characteristic Curve). The segmentation performance is evaluated by soft dice coefficient ($\mathcal{D}$) through multiple threshold level, set as (0.1, 0.3, 0.5, 0.7, 0.9). At each threshold level, the predicted probability map and soft ground-truth are binarized with the given threshold and then the dice metric (\cite{milletari2016v}) is computed. $\mathcal{D}$ scores are obtained as the averages of multiple thresholds.

\subsection{Experiment Results}

In the experiments, we first verify the effectiveness of the proposed DF-GT and T\&G Model by comparing them with corresponding SOTA methods in Section \ref{sec:eval_dfgt} and Section \ref{sec:eval_tg}, respectively. Then we verify the overall performance of DiFF with previous multi-rater learning methods in Section \ref{sec:diff}. Finally, we conduct the detailed ablation study to verify the effectiveness of each proposed component in Section \ref{section:ablation}.

\subsubsection{DF-GT vs SOTA Labels-fusion Methods}\label{sec:eval_dfgt}
\begin{figure}[h]
    \centering
    \includegraphics[width=\columnwidth]{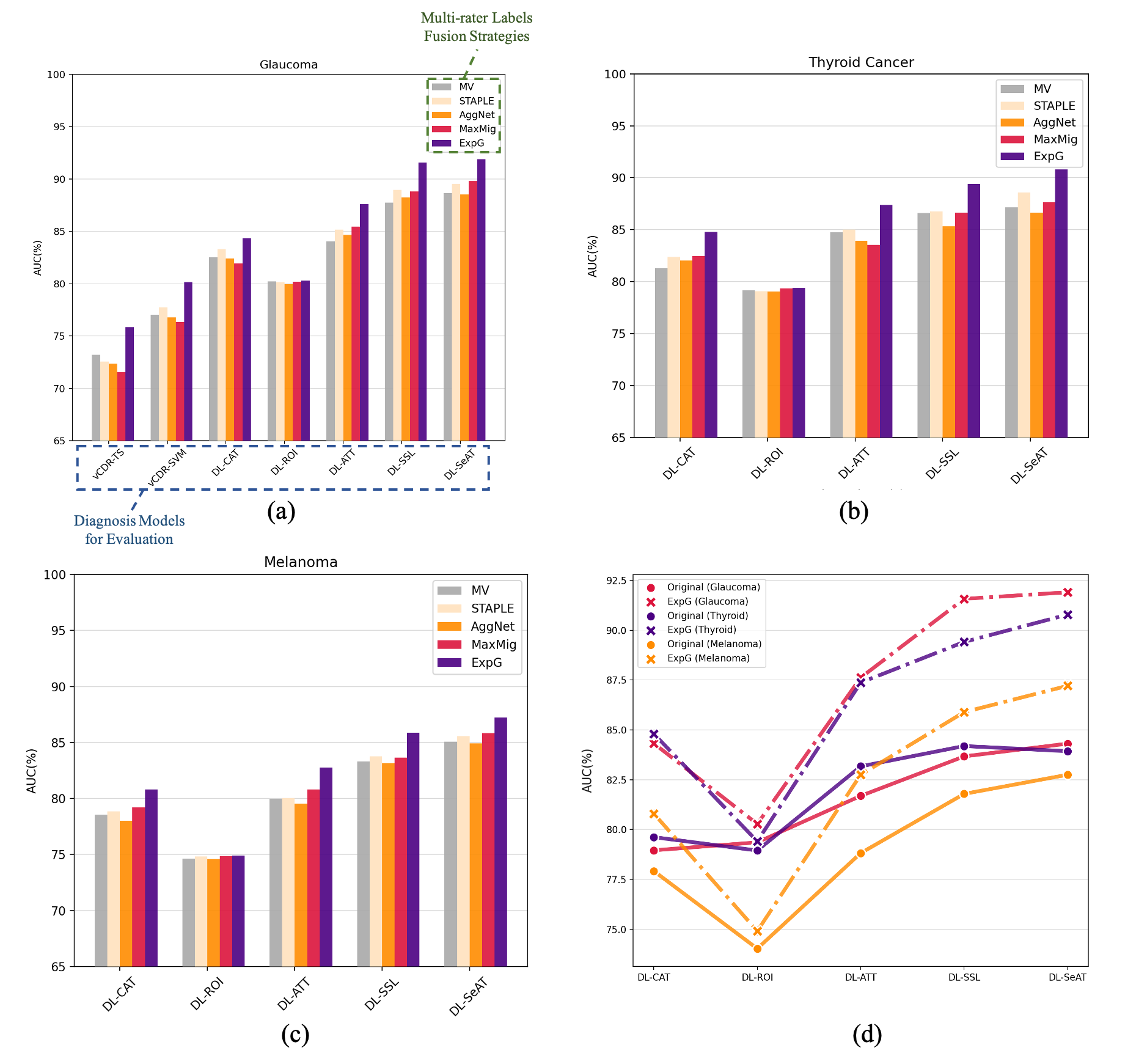}
    \caption{(a), (b), (c): Quantitative comparison of multi-rater labels fusion strategies: MV, STAPLE (\cite{warfield2004simultaneous}), AggNet (\cite{albarqouni2016aggnet}), MaxMig (\cite{cao2019max}), and proposed ExpG on glaucoma, thyroid cancer and melanoma diagnosis tasks respectively. We evaluate the diagnosis performance of them by various diagnosis models, including vCDR-TS (\cite{chandrika2013analysis}), vCDR-SVM (\cite{thangaraj2017glaucoma}), DL-CAT (\cite{d2021convit}), DL-ROI (\cite{bajwa2019two}), DL-ATT (\cite{li2019attention}), DL-SSL (\cite{wu2020leveraging}), and DL-SeAT (\cite{wu2022seatrans}). (d): Comparison of original high-frequency DF-GT (denoted as Original) and ExpG generated low-frequency DF-GT (denoted as ExpG) on different tasks.}
    \label{fig:barplot3}
\end{figure}

We propose DF-GT to learn the fusion of multi-rater segmentation labels in favor of the disease diagnosis. We compare DF-GT with SOTA multi-rater fusion strategies to verify that DF-GT contains more discriminative diagnosis features than the others. Specifically, we validate the diagnosis performance of the fused segmentation labels by various segmentation-assisted diagnosis approaches. In particular, the multi-rater labels are first fused to a unique ground-truth by the multi-rater fusion strategies. Then the unique ground-truth will be sent to a series of segmentation-assisted diagnosis models to evaluate its diagnosis contribution.
In the comparison, the multi-rater fusion strategies include majority vote (MV), STAPLE (\cite{warfield2004simultaneous}), AggNet (\cite{albarqouni2016aggnet}), and MaxMig (\cite{cao2019max}).
The segmentation-assisted diagnosis models we selected for the evaluation include the concatenation based method (DL-CAT) implemented by ResNet50 backbone (\cite{resnet}), ROI based method (DL-ROI) (\cite{bajwa2019two}), the Attention based method (DL-ATT) (\cite{li2019attention}), the Semi-supervised Learning based method (DL-SSL) (\cite{wu2020leveraging}), and the transformer based feature-fusion method (DL-SeAT) (\cite{wu2022seatrans}). Since directly calculating the vertical Cup-to-Disc Ratio (vCDR) on OD/OC is also a commonly used method in the clinical glaucoma diagnosis, we additionally use two CDR based diagnosis methods, vCDR-TS (\cite{chandrika2013analysis}) and vCDR-SVM (\cite{thangaraj2017glaucoma}), which implemented by thresholding and SVM on vCDR respectively, in glaucoma diagnosis task. The quantitative comparison measured by AUC (\%) on three different tasks is shown in Figure \ref{fig:barplot3}.

In Figure \ref{fig:barplot3}, 
we can see DF-GT outperforms the other multi-rater fusion methods on all of the segmentation-assisted diagnosis models. The improvement is more significant on DL-ATT, DL-SSL and DL-SeAT, which segmentation features are more efficiently used to facilitate the diagnosis. By contrast, the methods show similar performance on DL-ROI, which used the segmentation only for the localization. It demonstrates that DF-GT has a much higher diagnostic value than the other fused ground-truths. On vCDR based models of glaucoma diagnosis, DF-GT also 
presents the highest diagnosis performance, indicating it is generally effective for all kinds of diagnosis methods, not only the deep learning based ones.  

\begin{figure}[h]
    \centering
    \includegraphics[width=0.9\textwidth]{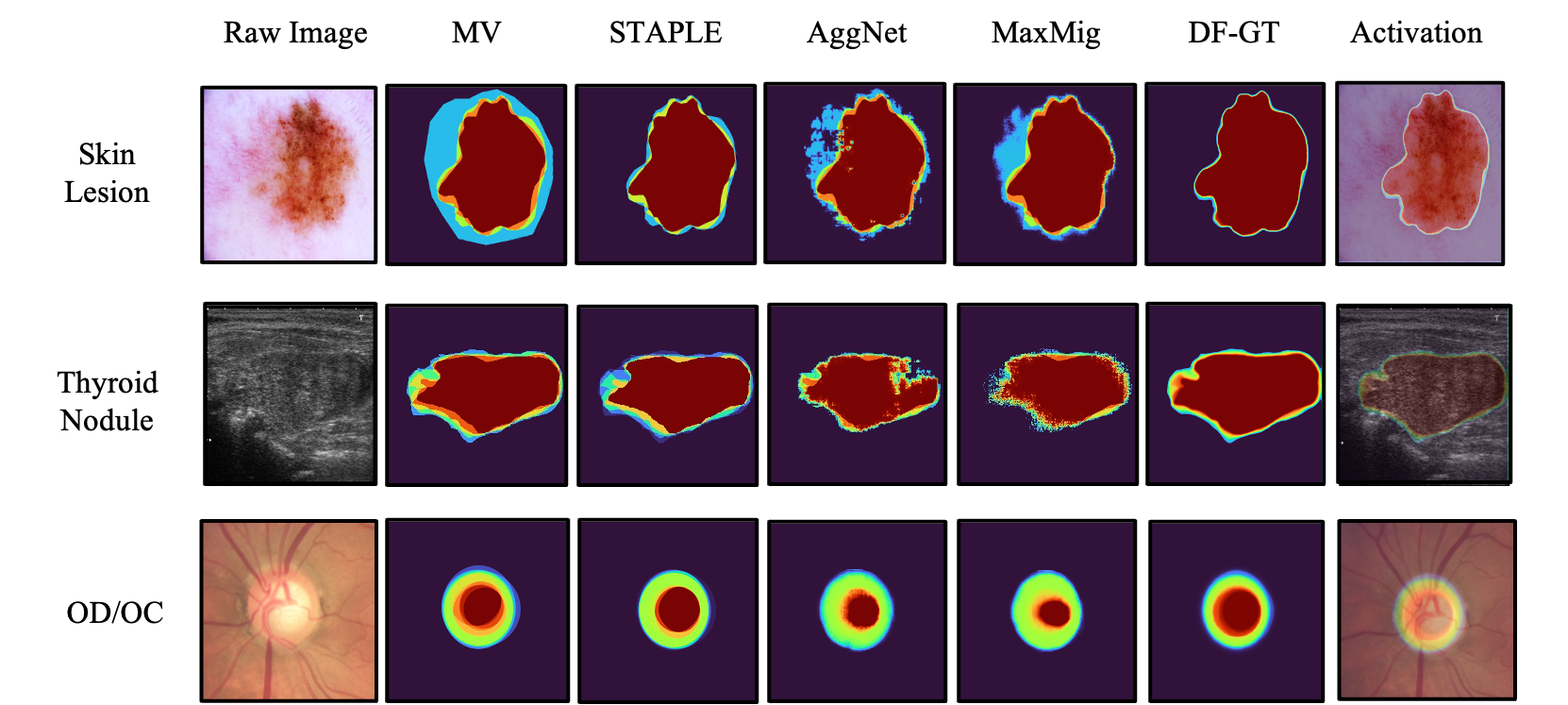}
    \caption{Visualized comparison of multi-rater fusion strategies. The examples are all positive cases, which means they are malignant melanoma, malignant thyroid nodule and glaucoma cases respectively. We can see DF-GT calibrate the uncertainty in a way to facilitate the diagnosis, e.g. calibrate the skin lesion with asymmetric structure (one side smooth and the other irregular), calibrate the thyroid nodule with irregular shape, and calibrate OD/OC with large vCDR.}
    \label{fig:df_vis}
\end{figure}

For a detailed comparison with the multi-rater fusion labels, we show the visualized comparison in Figure \ref{fig:df_vis}. We can see that MV and STAPLE highlight the regions where the multiple raters agreed and play down the uncertain regions. However, it has the chance that the minority is correct and the discriminative features are obliterated. Their diagnosis performance thus is constrained by the negative effects of these examples. AggNet and MaxMig depend on the segmentation prediction to fuse the labels. Thus they show inferior visualized and diagnosis performance on the tasks that the contours of the lesions/organs are more ambiguous (glaucoma and thyroid nodule). By contrast, the proposed DF-GT is able to pick and reserve the segmentation features that do good to the diagnosis thus achieves better performance. For example, on the skin lesion segmentation in Figure \ref{fig:df_vis}, we can see DF-GT uniquely stresses the asymmetrical and irregular structure of the lesion, which are the key signs of melanoma. On glaucoma diagnosis, DF-GT fuses the labels with larger vCDR, which is also a crucial biomarker indicating the glaucoma.

\subsubsection{T\&G Model vs SOTA Segmentation Methods}\label{sec:eval_tg}
Once we identified DF-GT as the ground-truth, the proposed T\&G Model can learn a better segmentation on it. We compare T\&G Model with previous SOTA segmentation methods on three different diseases. Since we find the segmentation models proposed for the specific diseases usually perform better than the general ones, we select the SOTA models proposed for each disease respectively for the comparison. On OD/OC segmentation, we compare with CENet (\cite{cenet}), AGNet (\cite{acnet}), BEAL (\cite{beal}), pOSAL (\cite{posal}) and ATTNet (\cite{ATTNet}). On thyroid nodule segmentation, we compare with TNSCNN (\cite{tnscnn}), CasCNN (\cite{cascnn}), MPCNN (\cite{mpcnn}), TRFE-Net (\cite{trfenet}) and TNSNet (\cite{tnsnet}). On skin lesion segmentation, we compare with DDN (\cite{ddn}), JaccNet (\cite{jaccnet}), Unver (\cite{unver}), FrCN (\cite{frcn}) and iFCN (\cite{ifcn}). Segmentation performance measured by Dice is shown in Table \ref{tab:sota}.


\begin{table}
\centering
\setlength\arrayrulewidth{0.1pt}
\caption{Comparing proposed T\&G Model with the other segmentation methods on various medical segmentation tasks. The segmentation performance is measured by Dice (\%).}
\resizebox{\columnwidth}{!}{%
\begin{tabular}{c|cc|c|c|c|c}
\hline
\multicolumn{3}{c|}{OD/OC} &\multicolumn{2}{c|}{Thyroid Nodule} &\multicolumn{2}{c}{Skin Lesions} \\ \hline
    Methods  &$\mathcal{D}_{disc}$ &$\mathcal{D}_{cup}$  & Methods &$\mathcal{D}$  &  Methods     &$\mathcal{D}$         \\ \hline
AGNet (\cite{acnet})    & 91.65      & 75.14     & TNSCNN (\cite{tnscnn}) & 80.77     & DDN (\cite{ddn}) &78.31   \\
CENet (\cite{cenet})     & 91.05      & 78.63   & CasCNN (\cite{cascnn})  &83.32    & JaccNet (\cite{jaccnet})    &83.25        \\
pOSAL (\cite{posal})   & 93.64      & 82.32      & MPCNN (\cite{mpcnn}) &83.64     &  Unver (\cite{unver})   &85.10     \\
BEAL (\cite{beal})     & 93.88      & 81.93      &  TRFE-Net (\cite{trfenet})     &84.83      & FrCN  (\cite{frcn})    &87.73        \\
ATNet (\cite{ATTNet})  & 95.78      & 84.65      &  TNSNet (\cite{tnsnet}) &85.43     & iFCN (\cite{ifcn})  &88.11      \\ \hline
T\&G & \textbf{96.30}   & \textbf{86.61}   & T\&G & \textbf{87.55} & T\&G & \textbf{89.48}   \\ \hline
\end{tabular}
}
\label{tab:sota}
\end{table}

In Table \ref{tab:sota}, we can see the proposed model consistently achieves superior segmentation performance compared with SOTA segmentation methods, which indicates T\&G Model can better predict DF-GT from raw images comparing with previous SOTA segmentation ones. The improvement is especially significant when the object is ambiguous. For example, proposed T\&G Model outperforms the previous SOTA by a remarkable 2.96\% on the optic cup, which is an tissue quite hard to be discriminated from the raw images. It indicates that T\&G Model can select and enhance the unique features that are conducive to the diagnosis, rather than only using the structural features on the raw images like the most other segmentation methods. As we have shown DF-GT is more beneficial to the diagnosis in the last section, this stronger segmentation capability of T\&G Model on DF-GT implies its predicted segmentation masks can also be more beneficial to the diagnosis, which is proved by the experimental results in the next section.

\subsubsection{DiFF vs SOTA Multi-rater Learning Methods}\label{sec:diff}
In this section, we show DiFF, a combination of DF-GT and T\&G Module can predict the segmentation maps that is more conducive to the diagnosis from the raw images. Toward that end, we conduct an overall comparison between DiFF and the previous multi-rater learning methods. The compared methods include the Bayesian based methods: sPU-Net (\cite{kohl2018probabilistic}) and PHiSeg (\cite{baumgartner2019phiseg}), the multi-head based method: WDNet (\cite{wdnet}), the Expectation-Maximization (EM) based methods: AggNet (\cite{albarqouni2016aggnet}) and MaxMig (\cite{cao2019max}), and calibrated segmentation method: MRNet (\cite{ji2021learning}). We quantitative evaluate the predicted segmentation results against each of the multi-rater labels and the self-fused labels. Self-fused labels refer to the weighted sum of the multi-rater labels by the self-predicted expertness, as shown in Eqn. (\ref{eqn:fusion}). Since in MRNet and WDNet, the expertness is user-provided, we evaluate their performance under majority vote following the setting of their papers. In sPU-Net and PHiSeg, we sample 16 times from the latent space following their setting and take the average as the result. The segmentation performance of the methods is measured by Dice. We evaluate the diagnosis capability of the predicted segmentation maps by Baseline-Dia. The quantitative comparison results on glaucoma, thyroid cancer and melanoma are shown in Table \ref{table:diff_glaucoma}, \ref{table:diff_thy} and \ref{table:diff_melanoma} respectively. 

Comparing the diagnosis capabilities of the methods, we can see DiFF consistently achieves superior performance, outperforms the second best method by 2.73\%, 2.51\% and 2.04\% AUC on glaucoma, thyroid cancer and melanoma diagnosis, respectively. This indicates that in the scenarios concerned with the final diagnostic effect, the proposed DiFF can provide more effective results. Comparing the segmentation performance on multi-rater labels, DiFF achieves top-3 on almost all of the raters, but its distribution over the raters is not always consistent with the other methods. For example, on glaucoma diagnosis, most the other methods prefer R1, R2 and R6 among the seven raters, while DiFF rather prefers R1, R2 and R4. This discrepancy may come from the criteria that methods used to 
evaluate the rater expertness. Most previous methods leverage the raw image prior to evaluate the rater expertness, which means that the annotations with a structure more similar to that observed on the raw image will be given higher confidence. By contrast, DiFF leverages the diagnosis as the standard to evaluate the raters. As shown in Figure \ref{fig:motivation}, R4 is more beneficial to the diagnosis than R6. Thus the inclination toward R4 of DiFF demonstrates that it is able to recognize the diagnosis-focused raters and predict the corresponding features directly from the raw image. It is also worth noting that DiFF may also show consistent inclinations toward the raters with the other methods. For example, on thyroid cancer diagnosis, all these methods incline to R2, R3 and R4. The different point is DiFF shows a heavier inclination and thus wins the highest diagnosis score. It seems to indicate that a more precise annotation of the structures on the raw images can indeed bring higher diagnosis performance in some cases. However, when the segmentation performance has reached a certain level and the boundary of the target object is really ambiguous, like many of the optic cups on fundus images, simply following the raw image structure will not be effective anymore for the diagnosis. In such cases, some of the raters are likely to annotate the objects with their prior knowledge of disease diagnosis. For example, when the contours of OD/OC are ambiguous, the rater may deliberately annotate OD/OC with larger vCDR if he/she suspects the case is positive (glaucoma). In these cases, DiFF is able to get the message and enhance these features through diagnosis-first principle.

\begin{table}[h]
\centering
\caption{The quantitative comparison between the proposed DiFF and SOTA multi-rater learning algorithms on \textit{OD/OC Segmentation \& Glaucoma Diagnosis}. We evaluate the segmentation capability of the predicted segmentation maps on multi-rater (R1-R7) annotations and self-fusion labels. We evaluate the diagnosis capability of the predicted segmentation maps on Baseline-Dia. Best result is shown in \textbf{bold}.}
\resizebox{\columnwidth}{!}{%
\begin{tabular}{c|cccccccc|c}
\hline
\multicolumn{9}{c}{OD/OC Segmentation \& Glaucoma Diagnosis} \\ \hline

\multirow{2}{*}{Methods} & R1   & R2  & R3 & R4  & R5  & R6 & R7 & Self & Diagnosis  \\ \cline{2-10} 
&$\mathcal{D}$ &$\mathcal{D}$   &$\mathcal{D}$   &$\mathcal{D}$ &$\mathcal{D}$   &$\mathcal{D}$   &$\mathcal{D}$ &$\mathcal{D}$   &$AUC$ \\ \hline
sPU-Net (\cite{kohl2018probabilistic})     &82.61 &81.87 &79.45 &84.12 & 82.73 & \textbf{83.55} & 76.47 & 85.22 & 84.15   \\
PHiSeg (\cite{baumgartner2019phiseg})    &80.36 &83.21 &79.83 &83.59 & 82.10 & 81.73 & \textbf{78.26} & 84.73 & 83.60   \\
WDNet (\cite{wdnet})      & 79.21 & 76.73 & 75.38    & 75.79 & 80.24 & 78.24 & 73.29 & 81.48 & 81.41  \\
AggNet (\cite{albarqouni2016aggnet})  & 81.45 &79.13 &81.55 &77.71 &83.52 &80.56 &75.45 &85.12 &83.83  \\ 
MaxMig (\cite{cao2019max})    &82.34 &80.76 &81.04 &83.12 & \textbf{85.67} & 82.90 & 77.82 &85.78 &84.43  \\ 
MRNet (\cite{ji2021learning})      &\textbf{84.89} &83.25 &\textbf{82.78} &78.55 &85.28 &79.40 &76.55 &86.25 &85.70   \\  \hline
DiFF  &84.27 &\textbf{85.17} &81.61 &\textbf{86.44} &84.08 &80.25 &76.74 &\textbf{86.61} &\textbf{88.13}  \\ \hline
\end{tabular}%
}
\label{table:diff_glaucoma}
\end{table}

\begin{table}[h]
\centering
\caption{The quantitative comparison between the proposed DiFF and SOTA multi-rater learning algorithms on \textit{Thyroid nodules Segmentation \& Thyroid cancer Diagnosis}. We evaluate the segmentation capability of the predicted segmentation maps on multi-rater (R1-R5) annotations and self-fusion labels. We evaluate the diagnosis capability of the predicted segmentation maps on Baseline-Dia. Best result is shown in \textbf{bold}.}
\resizebox{\columnwidth}{!}{%
\begin{tabular}{c|cccccc|c}
\hline
\multicolumn{8}{c}{Thyroid nodules Segmentation \& Thyroid cancer Diagnosis} \\ \hline

\multirow{2}{*}{Methods} & R1   & R2  & R3 & R4  & R5   & Self & Diagnosis\\ \cline{2-8} 
&$\mathcal{D}$   &$\mathcal{D}$ &$\mathcal{D}$   &$\mathcal{D}$   &$\mathcal{D}$ &$\mathcal{D}$   &$AUC$ \\ \hline
sPU-Net (\cite{kohl2018probabilistic})    &81.43 &82.54 &81.75 &85.43 & \textbf{83.76}  &84.29 &82.73 \\
PHiSeg (\cite{baumgartner2019phiseg})    &80.82 &80.37 &83.29 &84.20 & 81.45 & 83.67 & 84.68 \\
WDNet (\cite{wdnet})      &76.32 &77.43 &79.59 &78.45 & 77.21 &80.38 &80.64  \\
AggNet (\cite{albarqouni2016aggnet})  &79.37 &80.32 &84.79 &83.11 &81.04   &84.85 &83.38\\ 
MaxMig (\cite{cao2019max})    &\textbf{82.21} &81.04 &82.62 &83.54 &81.66 &85.10 &82.90   \\ 
MRNet (\cite{ji2021learning})      &80.92 &81.31 &84.25 &85.42 &82.06 &84.72 &85.24   \\  \hline
DiFF  &82.18 &\textbf{84.53} &\textbf{86.21} &\textbf{85.89} &83.27 &\textbf{87.55} &\textbf{87.75}   \\ \hline
\end{tabular}%
}
\label{table:diff_thy}
\end{table}

\begin{table}[h]
\centering
\caption{The quantitative comparison between the proposed DiFF and SOTA multi-rater learning algorithms on \textit{Skin lesions Segmentation \& Melanoma Diagnosis}. We evaluate the segmentation capability of the predicted segmentation maps on multi-rater (R1-R4) annotations and self-fusion labels. We evaluate the diagnosis capability of the predicted segmentation maps on Baseline-Dia. Best result is shown in \textbf{bold}.}
\resizebox{\columnwidth}{!}{%
\begin{tabular}{c|ccccc|c}
\hline
\multicolumn{7}{c}{Skin lesions Segmentation \& Melanoma Diagnosis} \\ \hline

\multirow{2}{*}{Methods} & R1   & R2  & R3 & R4  & Self & Diagnosis\\ \cline{2-7} 
&$\mathcal{D}$ &$\mathcal{D}$   &$\mathcal{D}$   &$\mathcal{D}$ &$\mathcal{D}$   &$AUC$ \\ \hline
sPU-Net (\cite{kohl2018probabilistic})    & 87.78 & 85.69 &81.63 &86.37 & 87.79 & 80.15   \\
PHiSeg (\cite{baumgartner2019phiseg})    & 86.26 & 84.21 &82.02 &85.71 & 87.50 & 79.43  \\
WDNet (\cite{wdnet})   & 79.64 & 78.35 &80.40 &78.29  & 81.20 & 76.66\\
AggNet (\cite{albarqouni2016aggnet})  & 83.98 & 81.12 &82.78 &84.79 & 85.85 & 78.52   \\ 
MaxMig (\cite{cao2019max}) & 87.53 &85.75  & \textbf{85.13} & 86.45 & 87.44 & 80.17    \\ 
MRNet (\cite{ji2021learning}) &87.09 &86.76 &84.71 &\textbf{86.59} &88.42 &80.32   \\  \hline
DiFF  &\textbf{88.81} &\textbf{87.67} &83.21 &84.70 &\textbf{89.48} &\textbf{82.06}   \\ \hline
\end{tabular}%
}
\label{table:diff_melanoma}
\end{table}

\subsubsection{Ablation Study}\label{section:ablation}
\textbf{DF-GT}
ExpG is proposed in the paper to improve generalization of DF-GT. In order to verify the effectiveness of ExpG, we compare the DF-GT before and after the application of ExpG. We use the same evaluation setting as that of multi-rater labels fusion strategies in Section \ref{sec:eval_dfgt}. 
The diagnosis performance measured by AUC (\%) is shown in Figure \ref{fig:barplot3} (d). 

As shown in Figure \ref{fig:barplot3} (d), comparing with original high-frequency DF-GT, ExpG generated DF-GT consistently achieves higher performance on different diagnosis methods, especially on stronger networks, like DL-ATT, DL-SSL and DL-SeAT. Concretely, ExpG generated DF-GT outperforms the counterpart by a 5.93\% AUC on DL-ATT, a 7.90\% AUC on DL-SSL, a 7.59\% AUC on DL-SeAT on glaucoma diagnosis. The similar conclusion can also be draw from the other tasks. It clearly shows that ExpG generated DF-GT is more robust for different diagnosis networks.

\textbf{T\&G Model} We propose T\&G Model to segment DF-GT from raw images. In order to verify the effectiveness of T\&G Model, we compare the segmentation performance of Baseline-Seg on DF-GT before and after the application of T\&G Model. The quantitative results on OD/OC segmentation are shown in Table \ref{tab:ab_tg} (a).
We can see that compared with segmentation baseline without T\&G Model, T\&G Model equipped one shows a significant improvement on DF-GT. That is because the segmentation network that absents the diagnosis prior knowledge is hard to capture the variance of DF-GT from the raw images. 
The proposed T\&G Model instills diagnosis knowledge to the segmentation network, thus improves the segmentation performance on diagnosis-first labels.



\begin{table}[h]
\setlength\arrayrulewidth{0.1pt}
\caption{Ablation study on T\&G Model.}
\begin{subtable}[t]{0.48\textwidth}
\flushleft
\resizebox{0.99\columnwidth}{!}{%
\begin{tabular}{c|cc|c|c}
\hline
\multirow{2}{*}{} & \multicolumn{2}{c|}{OD/OC} & Thyroid & Skin  \\ \cline{2-5} 
                  & $\mathcal{D}_{disc}$             & $\mathcal{D}_{cup}$           & $\mathcal{D}$       & $\mathcal{D}$     \\ \hline
w/o TG            & 91.10        & 76.92       & 81.15   & 82.15 \\ \hline
w TG              & \textbf{95.30}        & \textbf{86.61}       & \textbf{87.55}   & \textbf{89.48} \\ \hline
\end{tabular}%
}
\caption{\footnotesize Segmentation performance of baseline network before and after the application of T\&G Model.}
\label{Table:GoldDiag}
\end{subtable}
\hspace{\fill}
\begin{subtable}[t]{0.48\textwidth}
\resizebox{0.95\columnwidth}{!}{%
\begin{tabular}{llll|ll}
\hline
\multicolumn{4}{c|}{Connected Blocks}                                                                 & \multicolumn{2}{c}{Ave Dice}           \\ \hline
\multicolumn{1}{c|}{B1} & \multicolumn{1}{c|}{B2} & \multicolumn{1}{c|}{B3} & \multicolumn{1}{c|}{B4} & \multicolumn{1}{c|}{$\mathcal{D}_{disc}$} & \multicolumn{1}{c}{$\mathcal{D}_{cup}$} \\ \hline

\multicolumn{1}{c|}{\checkmark}   & \multicolumn{1}{c|}{\checkmark}   & \multicolumn{1}{l|}{}   &                         & \multicolumn{1}{l|}{93.36}   &82.78                        \\ \hline
\multicolumn{1}{l|}{}   & \multicolumn{1}{c|}{\checkmark}   & \multicolumn{1}{c|}{\checkmark}   &                         & \multicolumn{1}{l|}{95.04}   &85.84                        \\ \hline
\multicolumn{1}{l|}{}   & \multicolumn{1}{c|}{}   & \multicolumn{1}{c|}{\checkmark}   &\checkmark                     & \multicolumn{1}{c|}{91.52}   &80.11                        \\ \hline
\multicolumn{1}{c|}{\checkmark}   & \multicolumn{1}{c|}{\checkmark}   & \multicolumn{1}{c|}{\checkmark}   &                         & \multicolumn{1}{l|}{\textbf{95.30}}   &\textbf{86.61}                        \\ \hline
\multicolumn{1}{l|}{}   & \multicolumn{1}{c|}{\checkmark}   & \multicolumn{1}{c|}{\checkmark}   &\checkmark                     & \multicolumn{1}{c|}{93.58}   &84.72                        \\ \hline
\multicolumn{1}{c|}{\checkmark}   & \multicolumn{1}{c|}{\checkmark}   & \multicolumn{1}{c|}{\checkmark}   &\checkmark                      & \multicolumn{1}{c|}{94.79}   & 85.34                        \\ \hline
\end{tabular}
}
\caption{\footnotesize Segmentation performance of connecting different T\&G Residual Blocks on OD/OC segmentation.}
\label{tab:ab_tg}
\end{subtable}
\end{table}

The selection of the connected layers of T\&G Model will also effect the final segmentation performance. To find which layers should be connected by T\&G Model, the ablation study is performed in Table \ref{tab:ab_tg} (b). We sequentially apply the proposed T\&G Model on four different residual blocks of the segmentation encoder, diagnosis network and segmentation decoder. We denote the residual blocks from shallow to deep as B1 to B4 in the table. As shown in the table, applying T\&G Model on the shallow layers seems to be more efficient than applying on the deeper layers. This may because the diagnosis feature in B4 may be too unique to be leveraged for the segmentation. Segmentation network achieves the best performance when connecting B1, B2 and B3, but the effect of connecting the middle two blocks (B2 and B3) is the most obvious. This indicates that the diagnosis features and segmentation features of the first block may have some overlaps. In the paper, we connected B1, B2 and B3 when applying T\&G Model.

\section{Conclusion}
In this work, we proposed the idea to calibrate inter-observer segmentation uncertainty via diagnosis-first principle. In practice, we designed DiFF, a framework which is able to segment the masks to improve the diagnosis performance. This was achieved by the use of DF-GT, a fusion of multi-rater segmentation labels in favor of the diagnosis, as well as T\&G Model equipped segmentation network to learn from DF-GT. Extensive empirical experiments demonstrated the overall superior diagnosis performance of DiFF. In our future work, we will explore the relationship between the diagnosis-first segmentation features and the clinical biomarkers, to explain how neural network leverage these features to make the diagnosis decision, and explore the way our human to visualize and analysis these diagnosis-first segmentation features.

\section*{Acknowledgements}
This work was supported by the National Natural Science Foundation of China (82121003); Beijing Natural Science Foundation (Z190023); National Natural Science Foundation of China (NSFC) (62072190); and Program for Guangdong Introducing Innovative and Enterpreneurial Teams (2017ZT07X183).

\bibliographystyle{model2-names.bst}\biboptions{authoryear}
\bibliography{refs}
\clearpage

\end{document}